\documentclass[%
 reprint, showpacs,
 amsmath,amssymb,
 aps,
 prl,
]{revtex4-1}

\usepackage{graphicx}% Include figure files
\usepackage{dcolumn}% Align table columns on decimal point
\usepackage{bm}% bold math
\newcommand{\ve}[1]{{\bf{#1}}}
\newcommand{\ma}[1]{\underline{\underline{#1}}}

\begin{document}

\preprint{APS/123-QED}

\title{Scattering theory from artificial piezoelectric-like meta-atoms and molecules}%

\author{Leonid Goltcman and Yakir Hadad}%
 \email{hadady@eng.tau.ac.il}
\affiliation{School of Electrical Engineering, Tel-Aviv University, Ramat-Aviv, Tel-Aviv, Israel, 69978
}%

\date{\today}% It is always \today, today,
             %  but any date may be explicitly specified

\begin{abstract}
Inspired by the natural piezoelectric effect, we introduce hybrid-wave  electromechanical meta-atoms and meta-molecules that consist of coupled electrical and mechanical oscillators with similar resonance frequencies. We propose an analytical model for the linearized electromechanical scattering process, and explore its properties based on first principles. We demonstrate that by exploiting the linearized hybrid-wave interaction, one may enable functionalities that are forbidden otherwise, going beyond the limits of today's metamaterials.
As an example we show an electrically deep sub-wavelength dimer of  meta-atoms with extremely sensitive response to the direction-of-arrival of an impinging electromagnetic wave.
This  scheme of meta-atoms and molecules may open ways for metamaterials with a plethora of  exciting dynamics and phenomena that have not been studied before with potential technological implications in radio-frequencies and acoustics.

%As opposed to natural piezoelectricity and photoelasticity, these novel metamaterials can be designed to exhibit very strong coupling between electromagnetic waves and mechanical vibrations, and therefore open the door for a plethora of new exciting dynamics and phenomena that have not been studied before with potential technological implications in radio-frequencies and acoustics.
\end{abstract}

\pacs{43.35.+d, 77.84.-s, 41.20.Jb}% PACS, the Physics and Astronomy
                             % Classification Scheme.
%\keywords{Metamaterials, electromechanics, piezoelectricity}%Use showkeys class option if keyword
                              %display desired
\maketitle

%\tableofcontents

%\section{Introduction}
\emph{Introduction}.\textemdash The field of metamaterials, has gained a lot of attention in recent years for its potential in achieving effective material properties and functionalities that do not exist in nature  \cite{EnghetaBook, CapolinoBook}. 
%These effective materials consist of two typical scales, a microscopic - the meta-atom (unit-cell) that is small on the wavelength, and a macroscopic - the lattice that may be arbitrarily large. The effective material properties are determined typically by those two, as well as by the nature of the mutual coupling.
In the attempt to improve and enrich the performance of such effective materials, different types of meta-atoms, as well as lattice arrangements have been explored. These include electric or magnetic or combined \cite{Alu2009, Chen2012}, anisotropic \cite{Catrysse2013,VanOrden2010}, all-dielectric \cite{Ginn2012,Lin2014}, dynamic and time modulated \cite{Hadad2016,Fan2015}, nonlinear \cite{Chen2010,Sievenpiper2011}, nonreciprocal \cite{Hadad2010,Kodera2013,Mazor2014,Coulais2017}, active and non-foster \cite{Mirzaei2013, Long2016}, as well as mechanical and acoustical \cite{Coulais2017, Fleury2013, Nash2015, Paulose2015, Khanikaev2015} metamaterials.
Hybrid-physics metamaterials that combine electromagnetic and mechanical properties are developed alongside other approaches for creating real time reconfigurable, tuneable, as well as nonlinear devices \cite{Zheludev2016}. Utilizing thermal \cite{Tao2009,Ou2011}, electrostatic \cite{Chicherin2006,Hand2007,Zhu2011}, magnetic \cite{Valente2015}, and optical actuation \cite{Zhao2010,Lapine2012,Karvounis2015} one can deform, on-demand and in fast rates, the metamaterial structure in the microscale and hence have superior control over its effective properties.
Another type of man-made crystalline structures that involve hybrid-physics are the so called PhoXonic crystals \cite{Eichenfield2009,Eichenfield2009}. Due to the very different velocities of light and sound, in a lattice of optomechanical cavities, an infrared photon and a gigahertz phonon have similar wavelength, hence giving rise to simultaneous photonic and phononic Bragg resonance and strong photon-phonon interaction.  These and similar optomechanical structures have been extensively explored as nonlinear metamaterials \cite{Schmidt2013}, tuneable GHz resonators \cite{Pfeifer2016}, for quantum processing \cite{Schmidt2012,Peano2015}, as a mean for studying  many body dynamics \cite{Buks2002,Lifshitz2003}, and quantum many body dynamics \cite{Ludwig2013}, as well as for long range synchronization \cite{Shah2015}.

As opposed to previous work, here we introduce hybrid-physics electromechanical meta-atoms and meta-molecules that consist of coupled electrical and mechanical oscillators with \emph{similar} resonance frequencies, and that operate in a \emph{linearizable} regime. We study the electromechanical scattering from these structures through a simple electromechanical  response matrix, and explore its characteristics using first principles. We study small clusters and take advantage of the fact that an electrically small scatterer can be acoustically  large to get functionalities that are forbidden otherwise. As an examples we show an ultra-sensitive electrically-deep-subwavelength direction-of-arrival sensor for electromagnetic waves.

\begin{figure}[ht]
\centering
\includegraphics[width=0.45\textwidth]{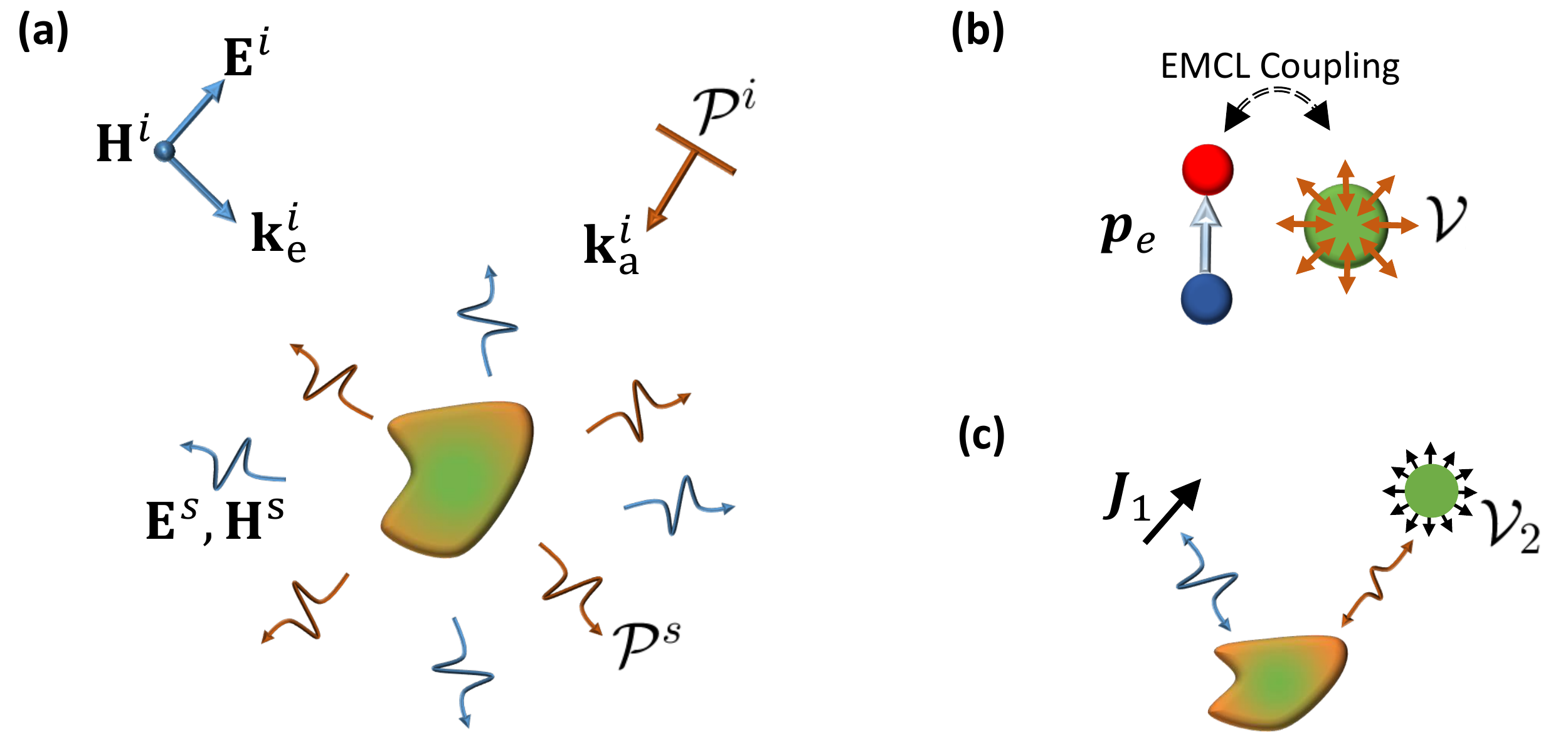}
\caption{(a) Illustration of the generalized hybrid-physics scattering process. An electromechanical meta-atom  can be excited by both electromagnetic or acoustic fields, and it generally scatters the two type of waves regardless on the excitation.  (b) If the resonators are electrically and acoustically small they compose a coupled system of electric dipole and acoustic monopole. Both radiate to the external ambient. (c) A configuration in which mutual action between the electric current source $\ve{J}_1$ and the acoustic pressure source ${\cal V}_2$ takes place via an EMCL meta-atom.}\label{Fig1}
\end{figure}

%\section{The electromechaical meta-atom}
\emph{The electromechanical meta-atom}.\textemdash Wave scattering typically occurs at the same `physics'. For instance we have electromagnetic scattering, acoustic scattering, elastic scattering, and so on.
%For instance, an electric scatterer transforms part of the energy of an impinging electromagnetic wave $\ve{E}^i$ into a scattered electromagnetic wave $\ve{E}^s$. Similarly, an acoustic scatterer partially transforms an impinging pressure wave ${\cal P}^i$ into a scattered pressure wave ${\cal P}^s$.
Here, however, we consider an hybrid-physics scattering.   An electromechanical (EMCL) scatterer  partially transforms an impinging electromagnetic $\ve{E}^i$ or acoustic  ${\cal P}^i$ waves into a mixture of  acoustic  and electromagnetic scattered waves $\ve{E}^s$ and ${\cal P}^s$, as illustrated in Fig.~\ref{Fig1}a. This type of scattering exists in natural  piezoelectric or photoelastic materials, however,  it may be better controllable and efficient using artificial materials that involve EMCL coupled resonators. Such artificial materials are composed of lattices of EMCL meta-atoms. The latter, are excited by and radiate EMCL fields.
%\subsection{EMCL response matrix - generalized ``polarizability''}
We define an EMCL field as a four element vector containing the three electric field components and the scalar pressure field $\ve{U}(\ve{r})=[E_x,E_y,E_z,{\cal P}]^T$.
When an EMCL field impinges an EMCL meta-atom, electric and acoustic sources are induced, as illustrated in Fig.~\ref{Fig1}b. Assuming that the meta-atom is small enough compare to the wavelength of light and sound,  the  induced sources are appropriately modelled by coupled electric dipole $\ve{p}_e$ and an acoustic monopole with volume $\cal V$ (so that its  \emph{volume velocity} is ${\cal U}=\dot{\cal V}$). These constitute the EMCL source  $\ve{S}=[p_{ex},p_{ey},p_{ez},{\cal V}]^T$.
Generally, the coupled EMCL problem is inherently nonlinear, however, in this work we restrict ourself to the class of problems that can be linearized under the weak signals assumption.    
%We further note that this type of modelling is valid also if the meta-atom is a bit larger, as long as at the operation frequency only these multipoles are close to resonance.
%We define the local field $\ve{U}^L(\ve{r})$, that is the field in the meta-atom location but in the absence of the meta-atom itself. Then, 
The induced source $\ve{S}$ is related to the impinging field $\ve{U}$ at the meta-atom location via the linear response matrix,
\begin{equation}\label{Generalized polarizability}
\ve{S}=\ma{\alpha}\ve{U}, \mbox{ with }
\ma{\alpha}=\left[  \begin{array}{cc}
                      \ma{\alpha}_{ee} & \ma{\alpha}_{ea} \\
                      \ma{\alpha}_{ae} & \ma{\alpha}_{aa} \\
                    \end{array}
 \right].
\end{equation}
The diagonal terms are the common response terms in the absence of EMCL coupling. Specifically,
\begin{equation}
\ma{\alpha}_{ee}=\left[\begin{array}{ccc}
                         \alpha_{ee}^{xx} & \alpha_{ee}^{xy} & \alpha_{ee}^{xz} \\
                         \alpha_{ee}^{yx} & \alpha_{ee}^{yy} & \alpha_{ee}^{yz} \\
                         \alpha_{ee}^{zx} & \alpha_{ee}^{zy} & \alpha_{ee}^{zz}
                       \end{array}
\right], \quad \ma{\alpha}_{aa}=\alpha_{aa}
\end{equation}
where $\ma{\alpha}_{ee}$ is the  electric polarizability that describes the induced dipolar moment due to an impinging electromagnetic field, and $\ma{\alpha}_{aa}$ gives the acoustic monopole volume induced by an impinging pressure field.
The off-diagonal, EMCL coupling, terms in Eq.~(\ref{Generalized polarizability}) read,
\begin{equation}
\ma{\alpha}_{ea}=\left[\alpha_{ea}^{x}, \alpha_{ea}^{y},\alpha_{ea}^{z} \right]^T, \quad   \ma{\alpha}_{ae}=\left[\alpha_{ae}^x,\alpha_{ae}^y,\alpha_{ae}^z \right].
\end{equation}
%\begin{equation}
%\ma{\alpha}_{ea}=\left[\begin{array}{c}
%                                          \alpha_{ea}^{x} \\
%                                          \alpha_{ea}^{y} \\
%                                          \alpha_{ea}^{z} \\
%                                       \end{array}
%\right], \quad   \ma{\alpha}_{ae}=\left[\alpha_{ae}^x,\alpha_{ae}^y,\alpha_{ae}^z \right].
%\end{equation}
These terms are responsible for the direct and reverse piezoelectric-like behaviour of the meta-atom.
Clearly, if the meta-atom exhibits no practical EMCL coupling then $\ma{\alpha}_{ae}=\ma{\alpha}_{ea}=0$, and if in addition it is only electric (acoustic) then $\ma{\alpha}_{aa}=0$ ($\ma{\alpha}_{ee}=0$).

%\subsection{The EMCL propagator}
%\emph{The EMCL propagator - }In the previous section we discussed the problem of EMCL source excitation by some impinging EMCL field. Here, we close the loop for the meta-atom model by considering the problem of radiation by an EMCL source $\ve{S}$.
Going to the opposite direction,  the field $\ve{U}(\ve{r})$ radiated by an induced source $\ve{S}$ on a meta-atom at $\ve{r}'$ is given by
%Once the EMCL meta-atom is exited it radiates an EMCL field.  The field $\ve{U}(\ve{r})$ that a source $\ve{S}$ located at $\ve{r}'$ radiates to an observer at $\ve{r}$ is given by 
the EMCL Green's function $\ve{U}(\ve{r})=\ma{G}(\ve{r},\ve{r}')\ve{S}$. Assuming  that there is no EMCL interaction in the ambient medium, $\ma{G}$ is block diagonal and reads
\begin{equation}
\ma{G}(\ve{r},\ve{r}')=\left[
               \begin{array}{cc}
                 \ma{G}_{e}(\ve{r},\ve{r}') & 0 \\
                 0 & G_{a}(\ve{r},\ve{r}') \\
               \end{array}
 \right]
\end{equation}
where $\ma{G}_e$ ($G_a$) is the electric dyadic (acoustic scalar) Green's function connecting $\ve{p}_e$ ($\cal V$) to $\ve{E}$ ($\cal P$).

%\subsection{Properties of the linear response matrix}
\emph{Fundamental constraints on $\ma{\alpha}.$}\textemdash The linear response matrix is subject to fundamental constraints due to reciprocity and energy conservation.
%\subsubsection{Reciprocity}
We begin with reciprocity.
%The polarizability of a bi-anisotropic small particles has to satisfy certain symmetries that are a consequence of reciprocity \cite{Kohenderink}. Here we follow similar argument but for the EMCL meta-atom.
Consider the hypothetical setup in Fig.~\ref{Fig1}c that contains an  electric current $\ve{J}_1$,  acoustic monopole with volume velocity ${\cal U}_2=\dot{\cal V}_2$, and an EMCL meta-atom.
In the absence of the meta-atom, the interaction between the two sources is obviously zero. However, in the presence of the EMCL meta-atom, the electric field radiated by the current source $\ve{J}_1$  impinges the meta-atom, and consequently gives rise to scattering of both electromagnetic and acoustic pressure waves. The latter, denoted here by ${\cal P}_1$, interacts with the acoustic source ${\cal U}_2$, implying that this time an action ${\cal A} \left[ \ve{J}_1\rightarrow{\cal U}_2\right]={\cal P}_1{\cal U}_2$ between the sources takes place. In the reciprocal scenario the acoustic source ${\cal U}_2$ acts on $\ve{J}_1$ through the scattered electromagnetic field $\ve{E}_2$,  ${\cal A} \left[ {\cal U}_2\rightarrow\ve{J}_1\right]=\ve{E}_2\cdot\ve{J}_1$. Since we deal with a linearized system the mutual action between the sources should be equal \cite{Kinsler},
\begin{equation}\label{Equal action - general}
{\cal A} \left[ \ve{J}_1\rightarrow{\cal U}_2\right] = {\cal A} \left[ {\cal U}_2\rightarrow\ve{J}_1\right].
\end{equation}
Expressing Eq.~(\ref{Equal action - general}) using the electromagnetic and acoustic Green's functions, we find
\begin{eqnarray}\label{Equal action}
&&{\cal U}_2 G_{a}(\ve{r}_2,\ve{r}_s)\ma{\alpha}_{ea}\ma{G}_{e}(\ve{r}_s,\ve{r}_1)\ve{J}_1 = \\ \nonumber &&\ve{J}_1^T\ma{G}_{e}(\ve{r}_1,\ve{r}_s)\ma{\alpha}_{ae}G_{a}(\ve{r}_s,\ve{r}_2){\cal U}_2.
\end{eqnarray}
Assuming that the medium is electromagnetically and acoustically reciprocal, $\ma{G}_{e}(\ve{r},\ve{r}')=\ma{G}_{e}^T(\ve{r}',\ve{r})$ \cite{Jackson}, $G_{a}(\ve{r},\ve{r}') = G_{a}(\ve{r}',\ve{r})$ \cite{Kinsler}. Then with Eq.~(\ref{Equal action}) we find
\begin{equation}\label{Pol Symetry rule}
\ma{\alpha}_{ea}=\ma{\alpha}_{ae}^T.
\end{equation}
This symmetry is  a manifestation of the principle of microscopic reversibility  \cite{Lewis,Onsager} applied to the linearized meta-atom system. 
%It should be compared with a similar result for the magnetoelectric polarizabilities of a bi-anisotropic scatterer $\ma{\alpha}_{em}=-\ma{\alpha}_{me}^T$, but with the `-' sign that is a consequence of the duality of Maxwell's equations \cite{Sersic2011}.

Next, we consider energy conservation. In the absence of material losses of any kind, the power that an impinging EMCL field $\ve{U}$ extracts  for the excitation of the induced source $\ve{S}$ on the meta-atom is equal to the total EMCL radiated power by the meta-atom.
The extracted EMCL power reads $P^{ext}=(\omega/2)\mbox{Im}\{\ve{U}^H\ma{\alpha}^H\ve{U} \}$ where superscript $H$ denotes the Hermitian transpose \cite{SM}. On the other hand, the total radiated power reads
$P^{rad}= \ve{U}^H\ma{\alpha}^H\ma{\chi}\ma{\alpha}\ve{U}$ with $\ma{\chi}=\mbox{diag}[\ma{I}_{3\times3}P^{rad}_e, P^{rad}_a]$. Where $\ma{I}_{3\times3}$ is the 3 by 3 unitary matrix, and  $P^{rad}_e$, $P^{rad}_a$ are the total power radiated by an electromagnetic dipole, and an acoustic monopole, both of unit amplitudes \cite{SM}. For a meta-atom embedded in a homogenous medium with permittivity and permeability $\epsilon$ and $\mu$, and with density $\rho_0$ we have $P^{rad}_e=\mu\omega^4/12\pi c_e$ \cite{Balanis} and $P^{rad}_a=\rho_0\omega^4/8\pi c_a$ \cite{Pierce}, where $c_e$ and $c_a$ are, respectively, the speed of light and sound in the medium. If the medium is more complex, the radiation terms should be corrected accordingly. For instance, for a meta-atom embedded in an ,electromagnetic transparent, acoustic hard-wall duct with cross section area $A_d$ that supports only plane wave we have $P_a^{rad}=\rho\omega^2c_a/4A_d$ \cite{Pierce}, while $P^{rad}_e$ unchanged.
By equating $P^{ext}=P^{rad}$ we find that $\ma{\alpha}$ is subject to
\begin{equation}\label{Energy Conservation}
\ma{\alpha}^H\ma{\chi}\ma{\alpha}=(\omega/4j)[\ma{\alpha}^H-\ma{\alpha}].
\end{equation}
This is a generalization of the optical theorem \cite{EnghetaBook, Jackson}.

\begin{figure}[ht]
\centering
\includegraphics[width=0.45\textwidth]{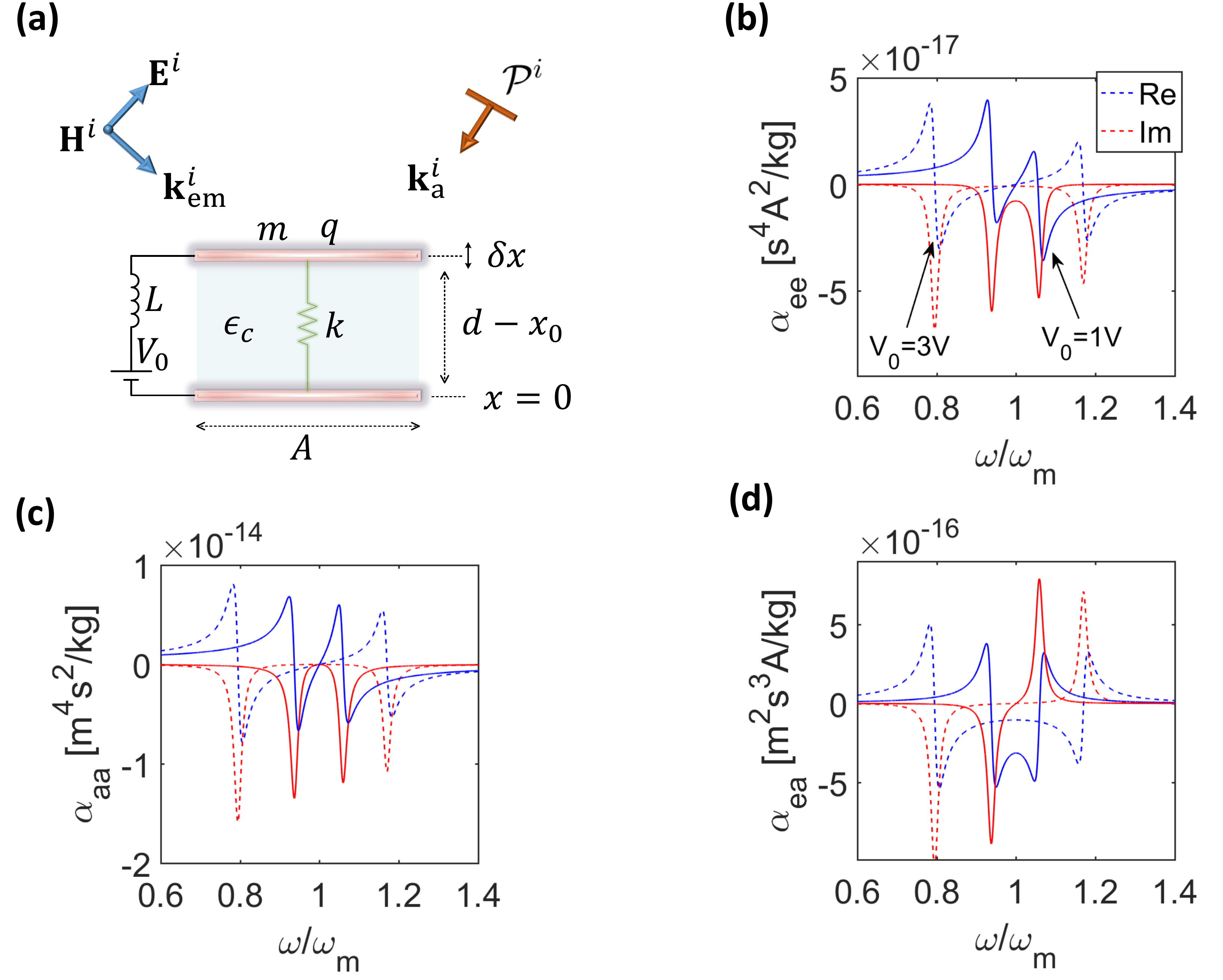}
\caption{(a) A parallel plate EMCL meta-atom, set at its operation point by a bias voltage $V_0$, and can be excited by electromagnetic or acoustic fields. Its EMCL small signal (linear) dispersion with frequency is given in (b-d). Blue (red) line denotes real (imaginary) part. Continuous  (dashed) line corresponds to biasing voltage $V_0=1V$  ($V_0=3V$). (b) The electric polarizabilty - $\alpha_{ee}$, (c) the acoustic response - $\alpha_{aa}$ - the induced acoustic monopole due to a local acoustic pressure field, (d) the EMCL  coupling terms - $\alpha_{ae}=\alpha_{ea}$ - induced electric dipole (acoustic monopole) due to a local acoustic pressure  (electric) field.   }\label{Fig2}
\end{figure}

%\section{The electromechanical meta-atom}
\emph{Schematic realization of EMCL meta-atom}.\textemdash
Consider a parallel plate capacitor with nominal capacitance $C_0$ loaded by an inductor $L$ to establish an electromagnetic resonance at frequency $\omega_e=1/\sqrt{LC_0}$. Simultaneously, each capacitor plate acts as a membrane that mechanically resonates at $\omega_m=\sqrt{k/m}$ where $m$, and $k$, are the membrane's effective mass and stiffness. We assume that the capacitor volume between the plates is acoustically closed, and thus it responds mechanically to external pressure changes. See Fig.~\ref{Fig2}a for illustration.
The system is set at equilibrium by applying a biasing voltage $V_0$, yielding to static charge accumulation, $q_0$ and $-q_0$,  and thereby  to a constant Coulomb attraction force between the plates. In the absence (presence)  of  the static biasing the spacing between the plates is $d$ ($d-x_0$). Neglecting edge effects, we define the nominal capacitance as $C_0=\epsilon_c A/(d-x_0)$ where $\epsilon_c$ is the permittivity between the plates, and $A$ is the plate area.

The meta-atom can be excited by either electromagnetic or acoustic wave as illustrated in Fig.~\ref{Fig2}a. Using the concept of effective length in antenna theory \cite{Balanis}, the impinging electromagnetic wave excitation is modelled by a lumped voltage source, $v(t)=l_{\mbox{\small eff}}E^i_x(t)$. Here, $E_x^i$ is the electric field component normal to the plates  and $l_{\mbox{\small eff}}$  is the effective length of the capacitor when viewed as an electrically small antenna. 
%Note that it is a property of the capacitor itself in the absence of the inductor, and moreover, since the distance between the plates is electrically small, $l_{\mbox{\small eff}}$ is only weakly dispersive. Similarly, the excitation by an acoustic wave can be modelled using a lumped force that pushes the membrane and is related to the pressure wave amplitude ${\cal P}^i$ via $f(t)=A_{\mbox{\small eff}}{\cal P}^i$ where $A_{\mbox{\small eff}}$ is the effective area of the plates \cite{SM}.  The lumped sources induce small signal charge and displacement fluctuations, $\delta q$ and $\delta x$, respectively.
Once excited, the meta-atom can be described effectively by an electric dipole
$\ve{p}_e=p_e\hat{x}$ with $p_e=l_{\mbox{\small eff}}\delta q$, coupled to an acoustic monopole with volume  ${\cal V}=A\delta x$  (volume velocity ${\cal U}=\dot{\cal V}=A\dot{\delta x} $)\cite{Kinsler}.
%Due to reciprocity (as shown below, assuming that $\delta q\ll q_0$ the system is practically linear, and thus reciprocal), since the meta-atom can be excited by electromagnetic and acoustic waves it can also emit both type of waves. Therefore, the electromechanical meta-atom  is in fact a coupled system of electric and acoustic small radiators. An electric  dipolar moment given by $\ve{p}_e=p_e\hat{x}$ with $p_e=l_{\mbox{\small eff}}\delta q$, coupled to an acoustic monopole pressure source with \emph{strength} of the spherical source (also termed volume velocity) $Q=A\dot{\delta x}$ \cite{Kinsler}.

The temporal dynamics is inherently nonlinear \cite{SM}. However, if the excitation is weak enough compared to static biasing so that $\delta q\ll q_0$, then the meta-atom response can be linearized around its equilibrium, %Eqs.~(\ref{Eq5})-(\ref{Eq6}) can be linearized. Yielding,
\begin{subequations}\label{Eq7}
\begin{eqnarray}
% \nonumber % Remove numbering (before each equation)
\ddot{\delta q}+ {2}{\tau_e^{-1}}\dot{\delta q}+\omega_e^2\delta q    &=& {L}^{-1} [v(t) + E_0\delta x], \\
\ddot{\delta x}+ {2}{\tau_m^{-1}}\dot{\delta x}+\omega_m^2\delta x &=&{m}^{-1}[f(t) + E_0\delta q].
\end{eqnarray}
\end{subequations}
%\begin{equation}\label{Eq7}
%\ddot{\delta q}+ \frac{2}{\tau_e}\dot{\delta q}+\omega_e^2\delta q    = \frac{1}{L} v(t) +\frac{1}{L} E_0\delta x,
%\end{equation}
%and
%\begin{equation}\label{Eq8}
%\ddot{\delta x}+\frac{2}{\tau_m}\dot{\delta x}+\omega_m^2\delta x =\frac{1}{m}f(t) + \frac{1}{m}E_0\delta q.
%\end{equation}
Here $\tau_e^{-1}$ and $\tau_m^{-1}$ are the electromagnetic and mechanical decay rates that include radiation as well as material damping,
$\omega_e$ and $\omega_m$ are defined earlier, 
%are the electromagnetic and mechanical  resonance frequencies in the absence of electromechanical coupling between the resonators,
and $E_0=-V_0/(d-x_0)$ is the static electric field between the capacitor plates.  The coupling terms in Eqs.~(\ref{Eq7}) have a clear physical meaning.
In Eq.~(\ref{Eq7}a) the small signal deflection $\delta x$ yields effectively an extra voltage source $E_0\delta x$, and in Eq.~(\ref{Eq7}b) the small signal charge $\delta q$ creates an extra force between the plates $E_0\delta q$.

The charge fluctuations ${\delta q}$  create an effective electric dipolar moment  ${p}_e=l_{\mbox{\small eff}}{\delta q}$, normal to the capacitor plates (along $\hat{x}$).
Moreover, the displacement fluctuations $\delta x$ give rise to an effective acoustic monopole source with volume oscillations amplitude ${\cal V}=A\delta x$ and volume velocity ${\cal U}=j\omega {\cal V}$
%
%Moreover, the meta-atom vibrates also mechanically and thereby radiates acoustic wave similar to the field of an acoustic monopole. Analogous to the dipolar moment of the electric dipole, the acoustic monopole is described uniquely using its volume velocity, given by ${\cal U}=j\omega {\cal V}$ where ${\cal V}=A {\delta x}$ is the volume oscillations amplitude.
(here and henceforth, time dependence $e^{j\omega t}$ is assumed and suppressed).
Finally, the system's linear response is expressed in the form of Eq.~(\ref{Generalized polarizability}), with
\begin{subequations}\label{Eq12}
\begin{eqnarray}
% \nonumber % Remove numbering (before each equation)
  \alpha_{ee} &=& ({l_{\mbox{\small eff}}^2}/{\Delta L})\left[\omega_m^2-\omega^2+{2j\omega}/{\tau_m} \right] \\
  \alpha_{aa} &=& ({A_{\mbox{\small eff}}^2}/{\Delta m})\left[\omega_e^2-\omega^2+{2j\omega}/{\tau_e} \right] \\
   \alpha_{ea} &=& \alpha_{ae}= {l_{\mbox{\small eff}}A_{\mbox{\small eff}}}E_0/{\Delta L}{m}
\end{eqnarray}
\end{subequations}
and where
\begin{equation}\label{Eq13}
\Delta = \left(\omega_e^2-\omega^2+\frac{2j\omega}{\tau_e} \right)\left(\omega_m^2-\omega^2+\frac{2j\omega}{\tau_m} \right)-\frac{E_0^2}{Lm}.
\end{equation}
In this example, the  meta-atom responds only to an $x$-polarized electric field, and therefore $\ma{\alpha}_{ee},\ma{\alpha}_{ea},\ma{\alpha}_{ae}$ are all scalars. Note the symmetry $\alpha_{ea}=\alpha_{ae}$ as dictated in Eq.~(\ref{Pol Symetry rule}) by reciprocity.
Moreover, assuming that the meta-atom  is lossless (namely, only radiation loss is allowed), using  Eq.~(\ref{Energy Conservation}) we find
\begin{eqnarray}\label{Energy Constrainst pp}
% \nonumber % Remove numbering (before each equation)
  {\omega}\Im\{\alpha_{ee}^{-1} \}/2 &=& P_e^{rad}+\left|{\alpha_{ae}}/{\alpha_{ee}} \right|^2P_a^{rad} \nonumber \\
  {\omega}\Im\{\alpha_{aa}^{-1} \}/2 &=& P_a^{rad}+\left|{\alpha_{ea}}/{\alpha_{aa}} \right|^2P_e^{rad}
\end{eqnarray}
and $\Im\{\alpha_{ee}^*\alpha_{ea} \}=\Im\{\alpha_{aa}^*\alpha_{ae} \}=\Im\{\alpha_{ee}^*\alpha_{aa} \}=0$. The latter three constraints are related to the mathematical structure of the linear response matrix, whereas the first two constraints given in Eq.~(\ref{Energy Constrainst pp}) can be solved to find the decay rates $\tau_e^{-1}$ and $\tau_m^{-1}$. In the absence of  static biasing, $V_0=0$, and therefore $\alpha_{ae}=\alpha_{ea}=0$, implying no EMCL coupling. In this case, the relations in Eq.~(\ref{Energy Constrainst pp}) are reduced to the conventional constraint of the polarizability of a small scatterer due to the optical theorem, and to its acoustic analog. By plugging Eqs.~(\ref{Eq12},\ref{Eq13}) into Eq.~(\ref{Energy Constrainst pp}), we get a nonlinear system that is solved for $\tau_e$, and $\tau_m$, yielding
\begin{equation}
\tau_e={\omega^2 L}/{l_e^2 P_e^{rad}}, \quad \tau_m={\omega^2 m}/{A_e^2 P_a^{rad}}.
\end{equation}

The decay rates are  proportional to the radiated power and hence the balance between $\tau_e$ and $\tau_m$ can be considerably tuned by appropriate  engineering of the meta-atom ambient medium. Since at a given frequency $\omega$, $\lambda_e=2\pi c_e/\omega\gg\lambda_a=2\pi c_a/\omega$,  a meta-atom that its typical size  is in the order of $\sim\lambda_a$ will be electrically deep subwavelength $\ll\lambda_e$. Therefore, typically, the electromagnetic radiation efficiency will be considerably smaller than its acoustic counterpart, implying that the electromagnetic resonance dominates since $\tau_e\gg\tau_m$. 
To make these rates comparable one may excite higher order acoustic moltipoles that are less efficient radiators, or reduce the ambient medium density. However the greatest control over the meta-atom decay rates will be obtained by placing it in an acoustic or electromagnetic duct or cavity with a suitably engineered local density of states.
%
%
%There are several alternatives to make these rates comparable, either by using higher order acoustic moltipoles that radiate less, or by reducing the density of the ambient medium which will lead to weaker acoustic coupling between meta-atoms in a cluster. However, a more controllable  approach would be to place the meta-atom in an acoustic or electromagnetic duct or cavity that, due to its modified  local density of states, affect the meta-atom acoustic and electromagnetic radiation  rates. 
The latter idea is demonstrated in Fig.\ref{Fig2}(b-d) where the  elements of the response matrix are plotted as function of frequency for the meta-atom in Fig.~\ref{Fig2}(a) with $\omega_e=\omega_m=2\pi\times10^6$rad/s, $A_{\mbox{eff}}=3.14\mu m^2$, $l_{\mbox{eff}}=10\mu$m, $m=4.2\mu$g,  $L=1\mu$H, that is embedded in an, electromagnetically transparent, hard-wall acoustic duct with cross section $A_d=5A_{\mbox{eff}}$ that  supports an acoustic plane wave only. The mechanical parameters are taken close to \cite{Purdy2012}. Here, $\tau_e\sim\tau_m$ and the system is in the strong coupling regime.  The coupling tunability  via the static biasing voltage is demonstrated with $V_0=1$V and $V_0=3$V.

\begin{figure}[ht]
\centering
\includegraphics[width=0.45\textwidth]{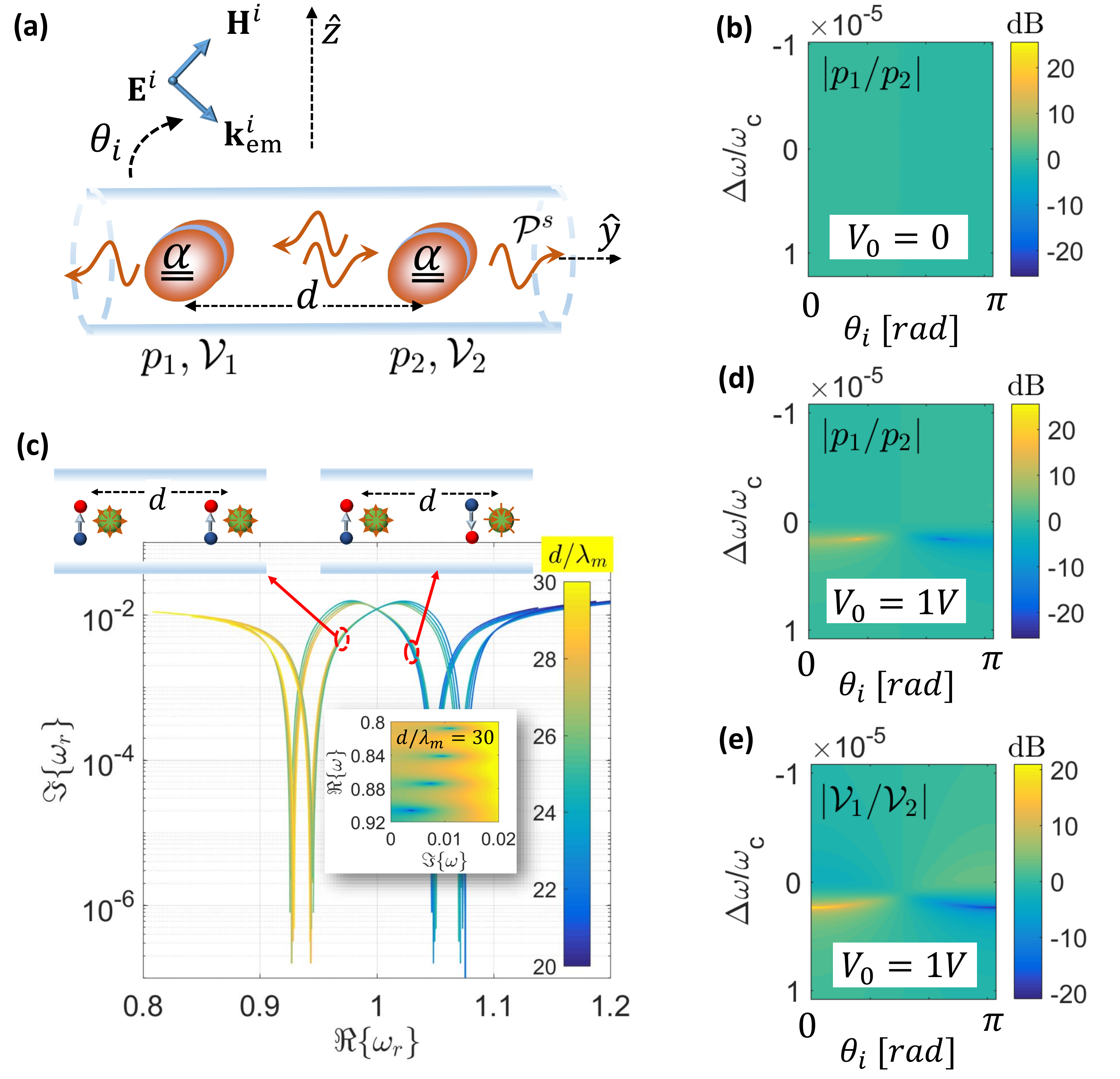}
\caption{(a) Illustration of an electrically deep subwavelength direction-of-arrival sensor for an electromagnetic wave.  (b) In the absence of EMCL coupling, $V_0=0$, only the bright mode can practically be excited and therefore the excitation of the two electric dipoles is practically identical for all $\theta_i$.  (c) Evolution of the complex resonance frequencies  as the spacing $d$ between the meta-atoms varies. There are two families of resonances that correspond to  bright and dark EMCL states (upper inset) (d) As the electromechanical coupling is turned on, $V_0=1V$, the electric dipole excitation highly depends on $\theta_i$, (e) and similarly the acoustic excitation response.}\label{Fig3}
\end{figure}

\emph{Highly-sensitive electrically small direction-of-arrival sensor}.\textemdash The EMCL meta-atoms discussed above can be used to design a piezoelectric-like meta-molecules with superior performance due to the joint acoustical and electromagnetic properties.
As an interesting example we design a system of  two meta-atoms inside a duct that is centered along the $\hat{y}$ axis, and with parameters as used  for Fig.\ref{Fig2}(b-d).  We excite the system only by an electromagnetic wave impinging at incidence angle $\theta_i$, so that $\ve{U}^i=[E_x^i,0]^T$ with $E_x^i=E_0\exp[-jk_e(\cos\theta_i\hat{y}-\sin\theta_i\hat{z})]$ ($k_e=\omega/c_e$). The electric field polarization $\hat{x}$ normal to the meta-atoms plates. See Fig.~\ref{Fig3}(a). We set the separation between the meta-atoms $d$  to be electrically deep sub-wavelength $d\ll\lambda_e$ while acoustically large $d\gg\lambda_a$.  The dynamics of the coupled system is given by
\begin{eqnarray}\label{DimerMeta}
% \nonumber % Remove numbering (before each equation)
  \ve{S}_1 &=& \ma{\alpha}\left[\ma{G}(\ve{r}_1,\ve{r}_2)\ve{S}_2 + \ve{U}^i(\ve{r}_1) \right]\nonumber \\
  \ve{S}_2 &=& \ma{\alpha}\left[\ma{G}(\ve{r}_2,\ve{r}_1)\ve{S}_1 + \ve{U}^i(\ve{r}_2) \right]
\end{eqnarray}
where $\ve{S}_1,\ve{S}_2$ are the EMCL excitation amplitudes of the  meta-atoms located at $\ve{r}_1=-d/2\hat{y}$ and $\ve{r}_2=d/2\hat{y}$. The Green's functions used here are given in \cite{SM}.
It is instructive to consider the corresponding eigenvalue problem  alongside with the excitation one. To find the eigenfrequencies we set $\ve{U}^i=0$, and require non-trivial solutions in Eq.(\ref{DimerMeta}). In the absence of the EMCL coupling, the system reduces to a simple coupled dipoles  that support two resonances, bright  and dark,  with eigenfrequencies nearly independent of  $d\ll\lambda_e$.  In this case, an impinging electromagnetic plane wave cannot practically excite the dark mode  but only its bright counterpart.  Fig.~\ref{Fig3}(b) shows the ratio $|p_1/p_2|$ (in log scale) as function of the incidence angle $\theta_i$ and the normalized frequency $\Delta\omega/\omega_c$ (where $\Delta\omega=\omega-\omega_c$), about the dark resonance $\omega_c$.  Even near the dark resonance  $\omega_c=0.985458\omega_m$,  there is neither a practical difference between the excitation amplitude  of the  dipoles, nor effect when varying the incidence angle $\theta_i$.  Here,  as opposed to a conventional direction-of-arrival sensor with two antennas separated by $d\sim\lambda_e/2$ \cite{Balanis}, the phase difference between the received signals in the two antennas $\sim k_e d$ is extremely small since $d\ll\lambda_e$.  However, we boost the small phase effect  by utilizing the presence of  EMCL coupling. Since the structure is acoustically large the number of eigenfrequencies is significantly larger, and their complex values strongly depend on $d$.
A typical complex-$\omega$ plane showing the resonance locations is given in the  inset inside Fig.~\ref{Fig3}(c) for $d/\lambda_m=30$ where $\lambda_m=2\pi c_a/\omega_m$. In Fig.~\ref{Fig3}(c)  the loci of several complex eigenfrequencies are plotted with $d$ as a parameter that its value is color encoded ($\Im\{\omega\}$ is in log scale to emphasize the resonance distinct locations). There are two families of eigenfrequencies that correspond to number of bright and dark modes (see upper inset).
As an example, we set $d=28\lambda_m\approx9.24$mm and find low loss dark resonance at $\omega_c=0.92846\omega_m$. When exciting the structure with an  electromagnetic plane wave at frequencies about the dark resonance, the dark resonance interplays with the bright resonance and  give rise to very strong variation of the excitation amplitudes as function of the incidence angle $\theta_i$ as shown in Fig.~\ref{Fig3}(d,e). This correlation can be used to estimate the direction of arrival. Moreover, in this scheme, a measurement of the excited acoustic field, as opposed to the  electromagnetic field that can be overwhelmed by the impinging wave, may increase detection sensitivity and noise fidelity.

\emph{Conclusions.}\textemdash Here we discussed a  paradigm for tuneable piezoelectric-like metamaterials and  utilized the fact that an electrically small scatterer is usually acoustically large, giving rise to hybrid-wave resonances that are unique to the electromechanical wave system. This concept with the other results in the paper may pave the way towards a plethora of new exciting dynamics and phenomena that have not been studied before and that may lead to technological implications in radio-frequencies and acoustics, such as ultra sensitive detectors, super-resolution imaging, and highly non-reciprocal devices.

%\begin{figure}[ht]
%\centering
%\includegraphics[width=0.45\textwidth]{Fig7}
%\caption{(a) Illustration of an electrically extremely subwavelength direction-of-arrival sensor for an electromagnetic wave. The device consists of two EMCL resonators with $d=28\lambda_m\ll\lambda_e$. An impinging electromagnetic wave hits the device at angle $\theta_i$. (b) In the absence of EMCL coupling, $V_0=0$, only the bright mode can be excites and the excitation of the two electric dipoles is practically identical for all $\theta_i$. (c) As the electromechanical coupling is turned on, $V_0=1V$, the electric dipole excitation highly depends on $\theta_i$, (d) and similarly the acoustic excitation response.   }\label{Fig4}
%\end{figure}

\begin{acknowledgments}
The Tel-Aviv University rector startup fund is acknowledged.
\end{acknowledgments}


\begin{thebibliography}{1}% Produces the bibliography via BibTeX.

\bibitem{EnghetaBook}
N. Engheta and R. W. Ziolkowski (editors). Metamaterials: Physics and Engineering Explorations.
Wiley-IEEE Press, 2006.

\bibitem{CapolinoBook}
F. Capolino (editor). Theory and Phenomena of Metamaterials. CRC Press, 2009.


%%%% General
%\bibitem{Eleftheriades2012}
%G. V. Eleftheriades, A. K. Iyer, and P. C. Kremer,  ``Planar negative refractive index media using periodically l-c loaded transmission lines,'' \emph{IEEE Trans. Microwave Theory and Techniques},
%{\bf50} (12), 2702–2712, (2002).
%
%\bibitem{Alu2005}
%A. Alu and N. Engheta,  ``Achieving transparency with plasmonic and metamaterial coatings,'' \emph{Phys.
%Rev. E}, {\bf72} 016623, (2005).
%
%\bibitem{Alu2007}
%A. Alu, M. G. Silveirinha, A. Salandrino, and N. Engheta,  ``Epsilon-near-zero metamaterials and
%electromagnetic sources: Tailoring the radiation phase pattern,'' \emph{Phys. Rev. B}, {\bf75}, 155410, (2007).


%%%% Magnetic
\bibitem{Alu2009}
A. Alu and N. Engheta,  ``The quest for magnetic plasmons at optical frequencies,'' \emph{Opt. Express},
{\bf17} (7), 5723–5730 (2009).

\bibitem{Chen2012}
W.-C. Chen, C. M. Bingham, K. M. Mak, N. W. Caira, and W. J. Padilla, ``Extremely subwavelength planar magnetic metamaterials,'' \emph{Phys. Rev. B} {\bf85}, 201104(R) (2012)


%%%% Anisotropic
\bibitem{Catrysse2013}
P. B. Catrysse and S. Fan. ``Routing of deep-subwavelength optical beams and images without reflection
and diffraction using infinitely anisotropic metamaterials,'' \emph{Adv. Mat.}, {\bf25} (2), 194–198
(2013).

\bibitem{VanOrden2010}
D. Van Orden, Y. Fainman, and V. Lomakin,  ``Twisted chains of resonant particles: optical polarization
control, waveguidance, and radiation,'' \emph{Opt. Lett.}, {\bf35} (15), 2579–2581 (2010).


%%%% All dielectric
\bibitem{Ginn2012}
James C. Ginn \emph{et al}, ``Realizing Optical Magnetism from Dielectric Metamaterials''
\emph{Phys. Rev. Lett.} {\bf108}, 097402  (2012)

\bibitem{Lin2014}
D. Lin, P. Fan, E. Hasman, and Mark L. Brongersma, ``Dielectric gradient metasurface optical
elements,'' \emph{Science}, {\bf345} (6194), 298–302 (2014).


%%%%% Dynamic (Time modulated)
%\bibitem{Hadad2015}
%Y. Hadad, D. L. Sounas, and A. Alu,  ``Space-time gradient metasurfaces,'' \emph{Phys. Rev. B}, {\bf92} 100304, (2015).

\bibitem{Hadad2016}
Y. Hadad, J. C. Soric, and A. Alu, ``Breaking temporal symmetries for emission and absorption,''
\emph{Proc. Nat. Acad. Sci.}, {\bf113} (13), 3471–3475 (2016).

\bibitem{Fan2015}
K. Fan and Willie J. Padilla, ``Dynamic electromagnetic metamaterials,'' \emph{Materials Today}, {\bf18}, (1) (2015).


%%%% Nonlinear

\bibitem{Chen2010}
P.-Y. Chen and A. Alu,  ``Optical nanoantenna arrays loaded with nonlinear materials,'' \emph{Phys. Rev.
B}, {\bf82},  235405 (2010).

\bibitem{Sievenpiper2011}
D. F. Sievenpiper, ``Nonlinear grounded metasurfaces for suppression of high-power pulsed rf currents,''
\emph{IEEE Antennas and Wireless Propagation Letters}, {\bf10}, 1516–1519 (2011).

%\bibitem{Wakatsuchi2013}
%H. Wakatsuchi, S. Kim, J. J. Rushton, and D. F. Sievenpiper,   ``Circuit-based nonlinear metasurface
%absorbers for high power surface currents,'' \emph{App. Phy. Lett.}, {\bf102} (21), (2013).



%%%% Nonreciprocal
\bibitem{Hadad2010}
Y. Hadad and Ben Z. Steinberg, ``Magnetized spiral chains of plasmonic ellipsoids for one-way
optical waveguides,'' \emph{Phys. Rev. Lett.}, {\bf105}, 233904 (2010).

\bibitem{Kodera2013}
T. Kodera, D. L. Sounas, and C. Caloz, ``Magnetless Nonreciprocal Metamaterial (MNM) Technology: Application to Microwave Components,'' \emph{IEEE Trans. Microwaves Theory and Thechniques}, {\bf61} (3) (2013)

\bibitem{Mazor2014}
Y. Mazor and Ben Z. Steinberg, ``Metaweaves: Sector-way nonreciprocal metasurfaces,'' \emph{Phys. Rev.
Lett.}, {\bf112}, 153901 (2014).

\bibitem{Coulais2017}
C. Coulais,	D. Sounas,	and A. Alu, ``Static non-reciprocity in mechanical metamaterials,'' \emph{Nature} {\bf542}, 461–464 (2017).


%%%% Non-foster
\bibitem{Mirzaei2013}
H. Mirzaei and G. V. Eleftheriades,  ``Realizing non-foster reactive elements using negative-group delay
networks,'' \emph{IEEE Trans. Microwaves Theory and Techniques}, {\bf61}, (12), 4322-4332 (2013).

\bibitem{Long2016}
J. Long and D. F. Sievenpiper,  ``Low-profile and low-dispersion artificial impedance surface in the
uhf band based on non-foster circuit loading,'' \emph{IEEE Trans. Ant. Prop.} , {\bf64} (7), 3003–3010 (2016).




%%%% Mechanical
\bibitem{Fleury2013}
R. Fleury and A. Alu, ``Extraordinary sound transmission through density-near-zero ultranarrow
channels,'' \emph{Phys. Rev. Lett.}, {\bf111}, 055501 (2013).


\bibitem{Nash2015}
L. M. Nash \emph{et al},  ``Topological mechanics of gyroscopic metamaterials,'' \emph{Proc. Nat. Acad. Sci.}, {\bf112} (47), 14495–14500 (2015).

\bibitem{Paulose2015}
J. Paulose, A. S. Meeussen, and V. Vitelli, ``Selective buckling via states of self-stress in topological
metamaterials,'' \emph{Proc. Nat. Acad. Sci.}, {\bf112} (25), 7639–7644 (2015).

\bibitem{Khanikaev2015}
A. B. Khanikaev, R. Fleury, S. H. Mousavi, and A. Alu,  ``Topologically robust sound propagation in
an angular-momentum-biased graphene-like resonator lattice,'' \emph{Nat.  Comm.}, {\bf6},  8260:1–
7 (2015).


%%Reconfigurable
\bibitem{Zheludev2016}
N. I. Zheludev, and E. Plum, ``Reconfigurable nanomechanical photonic metamaterials,'' \emph{Nat. Nanotech} {\bf11}, 16-22, DOI:10.1038 (2016)


%% Thermal actuation
\bibitem{Tao2009}
H. Tau \emph{et al}, ``Reconfigurable terahertz metamaterials,'' \emph{Phys. Rev. Lett.} {\bf103}, 147401 (2009)

\bibitem{Ou2011}
J. Ou, E. Plum, L. Jiang, and N. I. Zheludev, ``Reconfigurable photonic metamaterials,'' \emph{Nano Lett.}, {\bf 11}, 2142-2144 (2011).

%%Electrostatic actuation
\bibitem{Chicherin2006}
D. Chicherin \emph{et al}, ``MEMS-based high impedance surfaces for millimeter and submillimeter wave applications,'' \emph{Microw. Opt. Technol. Lett.}, {\bf48}, 2570-2573 (2006).

\bibitem{Hand2007}
T. Hand and S. Cummer, ``Characterization of tuneable metamaterial elements using MEMS switches,'' \emph{IEEE Antenn. Wireless Prop. Lett.}, {\bf6}, 401-404 (2007).

\bibitem{Zhu2011}
W. M. Zhu \emph{et al}, ``Switchable magnetic metamaterials using micromachining processes,'' \emph{Adv. Mater.} {\bf23}, 1792-1796 (2011)


%%Magnetic actuation
\bibitem{Valente2015}
J. Valente, J. Ou, E. Plum, I. J. Youngs, and N. I. Zheludev, ``Reconfiguring photonic metamaterials with currents and magnetic fields,'' \emph{App. Phys. Lett.} {\bf 106}, 111905 (2015).

%%Optical actuation
\bibitem{Zhao2010}
R. Zhao, P. Tassin, T. Koschny, C. M. Soukoulis, ``Optical forces in nanowire pairs and metamaterials,'' \emph{Opt. Express} {\bf18}, 25665-25676 (2010).

\bibitem{Lapine2012}
M. Lapine,	I. V. Shadrivov,	D. A. Powell, and Y. S. Kivshar, ``Magnetoelastic metamaterials,'' \emph{Nat. Mater.} {\bf11}, 30–33 (2012).

%\bibitem{Ginis2013}
%V. Ginis, P. Tassin, C. M. Soukoulis, and I. Veretennicoff, ``Enhancing optical gradient forces with metamaterials,'' \emph{Phys. Rev. Lett.}, {\bf 110}, 057401 (2013).

\bibitem{Karvounis2015}
A. Karvounis, J. Y. Ou, W. Wu, K. F. MacDonald, N. I. Zheludev, ``Nano-optomechanical nonlinear dielectric metamaterials,'' \emph{Appl. Phys. Lett.} {\bf107}, 191110 (2015)


%%Optomechanical arrays and Phoxonic crystals
\bibitem{Eichenfield2009}
M. Eichenfield, J. Chan, R. M. Camacho, K. J. Vahala, and O. Painter, ``Optomechanical crystals,'' \emph{Nature} {\bf462} (7269), 78–82 (2009).

\bibitem{Safavi-Naeini2010}
A. H. Safavi-Naeini and O. Painter, ``Design of optomechanical cavities and waveguides on a simultaneous bandgap phononic-photonic crystal slab,'' \emph{Opt. Express} {\bf18} (14), 14926–14943 (2010).


%%%Cavity optomechanics
%\bibitem{Aspelmeyer2014}
%M. Aspelmeyer, T. J. Kippenberg, and F. Marquardt, ``Cavity optomechanics,'' \emph{Rev. Mod. Phys.} {\bf86}, 1391  (2014)


%%Optomechanical metamaterials
\bibitem{Schmidt2013}
M. Schmidt, V. Peano, and F. Marquardt, ``Optomechanical metamaterials: Dirac polaritons, Gauge fields, and instabilities,'' \emph{arXiv}:1311.7095 (2013).

\bibitem{Pfeifer2016}
H. Pfeifer,T. Paraïso, L. Zang, and O. Painter, ``Design of tunable GHz-frequency optomechanical crystal resonators,'' \emph{Opt. Exp.} {\bf24}  (11), 11407-11419 (2016).

\bibitem{Schmidt2012}
M. Schmidt, M. Ludwig, and F. Marquardt, ``Optomechanical circuits for nanomechanical continuous variable quantum state processing,'' \emph{New J. Phys.} {\bf14} (12), 125005 (2012).



\bibitem{Peano2015}
V. Peano, C. Brendel, M. Schmidt, and F. Marquardt, ``Topological phases of sound and light,'' \emph{Phys. Rev. X} {\bf5} (3), 031011 (2015).



\bibitem{Buks2002}
E. Buks and M. L. Roukes, ``Electrically tunable collective response in a coupled micromechanical
array,'' \emph{Journal of Microelectromechanical Systems}, {\bf11} (6), 802–807  (2002).

\bibitem{Lifshitz2003}
R. Lifshitz and M. C. Cross, ``Response of parametrically driven nonlinear coupled oscillators with
application to micromechanical and nanomechanical resonator arrays,'' \emph{Phys. Rev. B}, {\bf67}, 134302 (2003).


\bibitem{Ludwig2013}
M. Ludwig and F. Marquardt, ``Quantum many-body dynamics in optomechanical arrays,'' \emph{Phys. Rev. Lett.} {\bf111} (7), 073603 (2013).


\bibitem{Shah2015}
S. Y. Shah, M. Zhang, R. Rand, and M. Lipson, ``Master-slave locking of optomechanical oscillators
over a long distance,'' \emph{Phys. Rev. Lett.}, {\bf114}, 113602  (2015).

\bibitem{Kinsler}
L. E. Kinsler and A. R. Frey, Fundamentals of Acoustics, 2nd Ed., John Wiely \& Sons (1962).


\bibitem{Jackson}
J. D. Jackson , \emph{Classical Electrodynamics}. Wiley: New York, 1998.


\bibitem{Lewis}
G. N. Lewis, ``A new principle of equilibrium,'' \emph{PNAS}, {\bf 11}, 179-183 (1921).

\bibitem{Onsager}
L. Onsager, ``Reciprocal relations in irreversible processes, I.,'' \emph{Phys. Rev.}, {\bf15} (37) (1931).


%\bibitem{Sersic2011}
%I. Sersic, C. Tuambilangana, T. Kampfrath, and A. F. Koenderink, ``Magnetoelectric point scattering theory for metamaterial scatterers,'', \emph{Phys. Rev. B}, {\bf83}, 245102 (2011).


\bibitem{SM}
Supplementary Material available online.

\bibitem{Balanis}
C. A. Balanis, Modern Antenna Handbook, John Wiely \& Sons  (2008).


\bibitem{Pierce}
A. D. Pierce, Acoustics - An introduction to its physical principles and applications, 3rd Ed., Acoustical Society of America (1991) [p. 161, p. 319].

\bibitem{Purdy2012}
P.-L. Yu, Y.P. Purdy and C.A. Regal, ``Control of material damping in high-q membrane microresonators,''
\emph{Phys. Rev. Lett,} {\bf108}:083603 (2012).

%\bibitem{DDA}
%B. T. Draine, and P. J. Flatau, ``Discrete-dipole approximation for scattering calculation,'' \emph{J. Opt. Soc. Am. A}, {\bf 11}, (4), 1491-99 (1994).




\end{thebibliography}
\end{document}

\preprint{APS/123-QED}

\title{Supplementary Material for ``Scattering theory from artificial piezoelectric-like meta-atoms and molecules''}%

\author{Leonid Goltcman and Yakir Hadad}%
 \email{hadady@eng.tau.ac.il}
\affiliation{School of Electrical Engineering, Tel-Aviv University, Ramat-Aviv, Tel-Aviv, Israel, 69978
}%

\date{\today}% It is always \today, today,
             %  but any date may be explicitly specified

\begin{abstract}
While the main text is written to be basically self-contained, there are some minor gaps that are less critical for the clear understanding of the message but are important for the reproduction of the paper results. These few items are described below  for the sake of completeness.
\end{abstract}

\maketitle

\onecolumngrid

\tableofcontents

\renewcommand{\thefigure}{S\arabic{figure}}
\setcounter{figure}{0}
\renewcommand{\theequation}{S\arabic{equation}}  

\section{Derivation of the extracted electromechanical power}
The power that an electromechanical wave $\ve{U}=[\ve{E},{\cal P}]^T$ extracts to excite an electromechanical meta-atom with induced source $\ve{S}=[\ve{p}_e,{\cal V}]^T$
is given by 

\begin{equation}
P^{ext}=\frac{1}{2}\Re\{\ve{J}^*\cdot\ve{E} + {\cal U}^*{\cal P} \}
\end{equation}
where $\ve{J}=j\omega\ve{p}_e$ is the current source associated with the induced electric dipole on the scatterer, and ${\cal U}=j\omega{\cal V}$ is the volume velocity associated with the monopole volume amplitude $\cal V$. Then, we have, 
\begin{equation}
P^{ext}=\frac{\omega}{2}\Im\{\ve{p}_e^*\cdot\ve{E} + {\cal V}^*{\cal P} \}=\frac{\omega}{2}\ve{S}^{H}\ve{U}.
\end{equation} 
However, since $\ve{S}=\ma{\alpha}\ve{U}$ (Eq. (1) in the main text), we can write
\begin{equation}
P^{ext}=\frac{\omega}{2}\Im\{(\ma{\alpha}\ve{U})^{H}\ve{U}\}=\frac{\omega}{2}\Im\{\ve{U}^H\ma{\alpha}^H\ve{U}\}.
\end{equation}
This completes the derivation of the extracted power given in the main text.

\section{Derivation of the total radiated power from an EMCL meta-atom}
The total radiated power from an EMCL source $\ve{S}$ in electromagnetically as well as acoustically homogeneous medium reads
\begin{equation}
P^{rad}=\frac{\rho\omega^4}{8\pi c_a}{\cal V}^*{\cal V} + \frac{\mu\omega^4}{12\pi c_e}\ve{p}_e^*\cdot\ve{p}_e.
\end{equation}
For a general medium, however, although the coefficients in the equation above will be changed, the square dependence on ${\cal V}^*{\cal V}$ and $\ve{p}_e^*\cdot\ve{p}_e$ will be maintained. Therefore, we define (as in the main text) $P_a^{rad}$ and $P_e^{rad}$ as the total radiated power for acoustic monopole and electromagnetic dipole sources with unit amplitude, namely, with $|{\cal V}|=1$ and $|\ve{p}_e|=1$. 
Then, by defining 
\begin{equation}
\ma{\chi}=\mbox{diag}[P_e^{rad},P_e^{rad},P_e^{rad},P_a^{rad}] = \mbox{diag}[P_e^{rad}\ma{I}_{3\times3}, P_a^{rad}]. 
\end{equation}
we obtain the expression given in the main text for the radiated power, 
\begin{equation}
P^{rad}=\ve{U}^H\ma{\alpha}^H\ma{\chi}\ma{\alpha}\ve{U}.
\end{equation}

\section{Lumped sources model for the parallel plates meta-atom}
For the parallel plate meta-atom discussed in Fig. 2(a) of the main text the impinging electric and pressure fields are modelled using lumped sources. The sources excitation scheme is shown in Fig.~\ref{FigS1} below.

\begin{figure}[ht]
\centering
\includegraphics[width=0.45\textwidth]{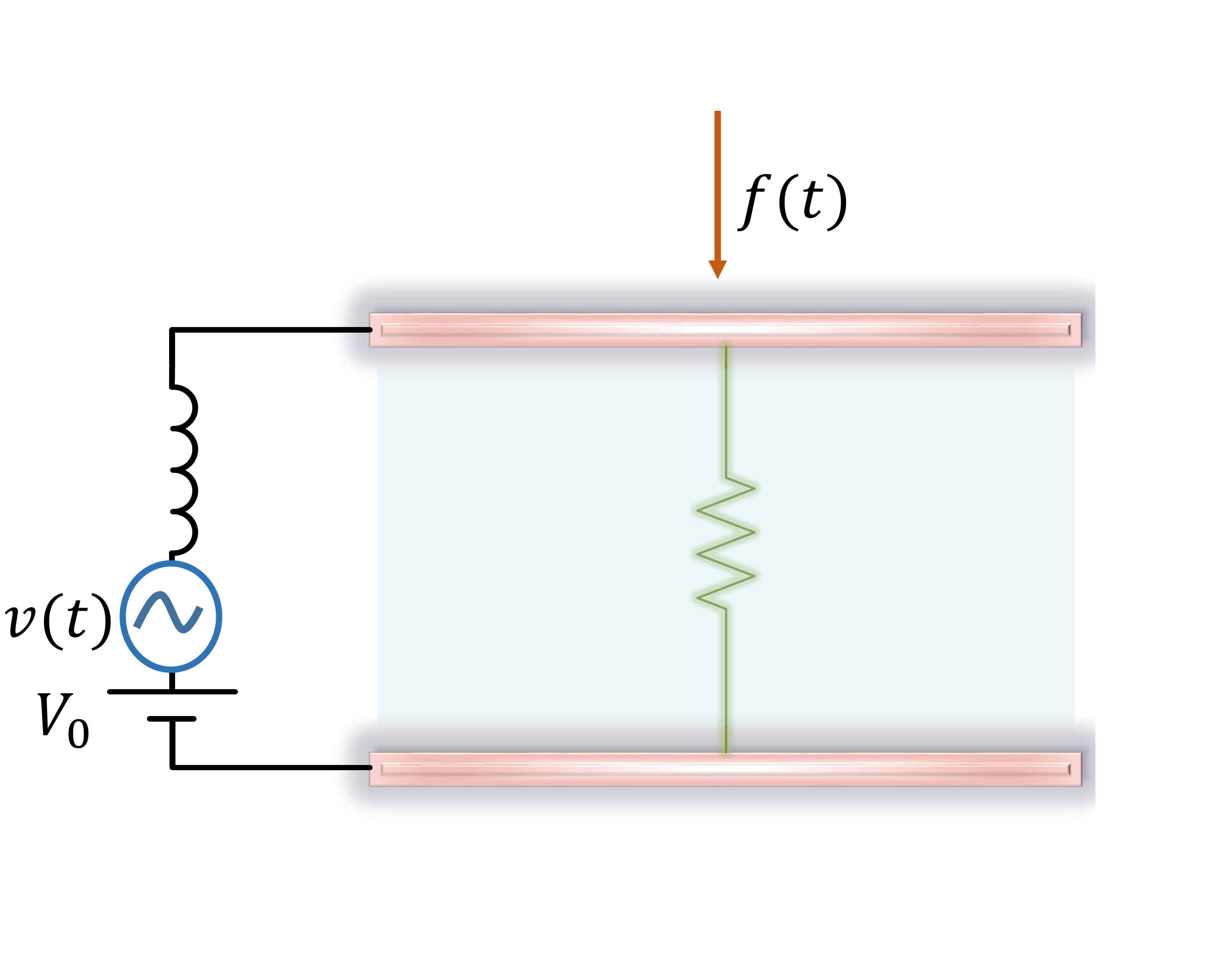}
\caption{Lumped excitation model}\label{FigS1}
\end{figure}

\section{Derivation of the parallel plates meta-atom dynamics}
The meta-atom dynamics it captured by the Lagrangian
\begin{equation}\label{Eq1}
{\cal L} = \frac{1}{2}m\dot{x}^2-\frac{1}{2}kx^2 + \frac{1}{2}L\dot{q}^2 - \frac{1}{2}\frac{q^2}{C(x)},
\end{equation}
with source ${\cal L}_s$ and dissipation $\cal F$ functionals
\begin{equation}\label{Eq2}
{\cal L}_s=qV+xf, \quad {\cal F}=\frac{1}{2}R\dot{q}^2 + \frac{1}{2}B\dot{x}^2
\end{equation}
where $V=V_0+v(t)$, $x=-x_0+\delta x$ ($x_0$ is taken such that $x_0>0$), and $q=q_0+\delta q$ are the plate's deflection and charge accumulation, respectively.
%As discussed above, $x_0$, and $q_0$ are due to the DC bias $V_0$ whereas the small signal fluctuations, $\delta x$ and $\delta q$, are due to the small signal excitation voltage $v(t)$ and force $f(t)$.
The dissipation term $\cal F$  includes two loss mechanisms, electromagnetic and mechanical,  represented, respectively, by resistance $R$ and viscosity $B$, where both consist of radiation as well as material loss channels.
%Using energy conservation $R_r=k_e^2\eta_0l_{\mbox{eff}}^2/6\pi$ where $k_e=\omega/c_e$ is the electromagnetic free space wavenumber and $c_e$  is the speed of light in the  ambient medium. $\eta=\sqrt{\mu_0/\epsilon_0}$  is the electromagnetic wave impedance.
%Similarly, $B=B_r+B_\Omega$ represent mechanical loss that consists of  material, $B_{\Omega}$, and  acoustic radiation loss $B_r$.
%Using energy conservation $B_r=\rho c_a k_a^2 A_{\mbox{\small eff}}^2/4\pi$ where $k_a=\omega/c_a$ is the free space sound wavenumber and $c_a$ is the speed of sound in the  ambient medium. $\rho$ is the medium acoustic density.
Using Eqs.(\ref{Eq1})-(\ref{Eq2}) in the Lagrange equations we obtain the relations between the static quantities, $x_0,q_0,V_0$, as well as  relations between the small signal terms $\delta x, \delta q, v(t), f(t)$, that represent the temporal dynamics.
For the static terms we find $q_0=C_0V_0,\quad kx_0=q_0^2/2\epsilon A$
%\begin{equation}\label{Eq3}
%q_0=C_0V_0,\quad kx_0=\frac{1}{2}\frac{q_0^2}{\epsilon A},
%\end{equation}
which  are nothing but the voltage-charge relation on the capacitor, and equality of forces between the spring repulsion due to contraction of $x_0$ and the Coulomb attraction between the plates of capacitor charged with $q_0$. Assuming that $x_0\ll d$ we may approximate $x_0\approx \epsilon A V_0/2kd^2$ and $q_0\approx \epsilon A V_0/d$.

For the dynamic terms we find the following nonlinear system
\begin{equation}
\ddot{\delta q}+2\tau_e^{-1}\dot{\delta q}+\omega_e^2\left(q_0\frac{\delta x}{d-x_0} + \delta q +\frac{\delta x}{d-x_0}\delta q \right)  = \frac{1}{L}v(t),
\end{equation}
and
\begin{equation}
\ddot{\delta x}+2\tau_m^{-1}\dot{\delta x} +\omega_m^2\delta x +\frac{1}{m}\frac{V_0}{d-x_0}\delta q +\frac{1}{2}\frac{\delta q^2}{C_0}\frac{1}{m(d-x_0)}=\frac{1}{m} f(t).
\end{equation}
Here $\tau_e^{-1}=R/2L$ and $\tau_m^{-1}=B/2m$ are the electromagnetic and mechanical damping rates, $\omega_e$ and $\omega_m$  are the electromagnetic and mechanical  resonance frequencies in the absence of electromechanical coupling between the resonators, and
$E_0=-V_0/(d-x_0)$ is the average electric field between the capacitor plates due to the static biasing.  
Eq.~(9) in the main text is derived from the nonlinear system above under the assumption that $\delta q\ll q_0$ and $\delta x\ll x_0$.

\section{Electromagnetic and acoustic Green's function used }
The electromagnetic Green's function in electromagnetically homogeneous medium with permittivity and permeability $\epsilon$ and $\mu$ is used in the calculation of the dimer response in Eq. (14) of the main text. It reads
\begin{equation}
\ma{G}_e(\ve{r},\ve{r}')=\frac{e^{-jk_er}}{4\pi\epsilon}\left\{\ma{A}\frac{k_e^2}{r} + [3\ma{B}-\ma{I}_{3\times3}]\left( \frac{1}{r^3}+\frac{jk_e}{r^2} \right) \right\}
\end{equation}
with
\begin{equation}
\ma{A}=\left[
\begin{array}{ccc}
  n_y^2+n_z^2 & -n_xn_y & -n_xn_z \\
  -n_xn_y & n_x^2+n_z^2 & -n_zn_y \\
  -n_zn_x & -n_zn_y & n_x^2+n_y^2 
\end{array}\right]
\end{equation}
and
\begin{equation}
\ma{B}=\left[
\begin{array}{ccc}
  n_x^2 & n_xn_y & n_xn_z \\
  n_xn_y & n_y^2 & n_zn_y \\
  n_zn_x & n_zn_y & n_z^2
\end{array}\right]
\end{equation}
where $\ma{I}_{3\times3}$ is the 3 by 3 unitary matrix, $r=|\ve{r}-\ve{r}'|$ is the distance between the source and observer points, and $\hat{n}=(n_x,n_y,n_z)=(\ve{r}-\ve{r}')/r$ is the unit vector pointing between the source and the observer. Moreover, $k_e=\omega/c_e$ where $c_e$ is the speed of light. For the dimer problem considered in the last part of the paper we had $\hat{n}=(0,\pm1,0)$ and $\ve{p}_{1,2}=p_{1,2}\hat{x}$.

The acoustic Green's function we used in that example was the Green's function of small acoustic monopole source in a duct supporting a plane wave only at the operation frequency (fundamental mode of the duct). In this case, since the driving frequency is below the cutoff of the higher order modes, the latter decays away from the source. Therefore, sufficiently far from the source (in terms of acoustic wavelength $d\gg\lambda_a$), we can asymptotically approximate the acoustic Green's function by
\begin{equation}
G_a\sim\frac{-j\omega\rho c_a}{2A_d}e^{-jk_a r}
\end{equation}
where here $r$ is the distance between the source and the observer along the duct axis. In the main text this is the $\hat{y}$ axis and the distance between the two meta-atoms ranged between $d=20\lambda_m$ and $d=30\lambda_m$, in any case satisfying the asymptotic assumption requirement. Moreover $k_a=\omega/c_a$, where $c_a$ is the speed of sound in the duct and $A_d$ is the duct cross section area.